\UseRawInputEncoding
\documentclass[aps,prl,floatfix,superscriptaddress,twocolumn,footinbib]{revtex4-1}
\usepackage{amssymb}
\usepackage{amsmath}
\usepackage{amsfonts}
\usepackage{appendix}
\usepackage{bm}
\usepackage{graphicx}
\usepackage{epsfig}
\usepackage{epstopdf}
\usepackage{balance}
\usepackage[dvipsnames]{xcolor}
\usepackage{calc}
\usepackage{natbib}
\usepackage[colorlinks,
            linkcolor=blue,
            anchorcolor=blue,
            citecolor=blue,
            urlcolor=blue]{hyperref}
\usepackage{lipsum}

\begin{document}
\title{ Doubled Moir\'{e} Flat Bands in Double-twisted Few Layer Graphite}
\author{Zhen Ma}
\affiliation{School of Physics and Wuhan National High Magnetic Field Center,
Huazhong University of Science and Technology, Wuhan 430074,  China}
\author{Shuai Li}
\affiliation{School of Physics and Wuhan National High Magnetic Field Center,
Huazhong University of Science and Technology, Wuhan 430074,  China}
\author{Ming Lu}
\affiliation{Beijing Academy of Quantum Information Sciences, Beijing 100193, China}
\affiliation{International Center for Quantum Materials, School of Physics, Peking University, Beijing 100871, China }
\author{Dong-Hui Xu}
\affiliation{Department of Physics, Hubei University, Wuhan 430062, China}
\author{Jin-Hua Gao}
\email{jinhua@hust.edu.cn}
\affiliation{School of Physics and Wuhan National High Magnetic Field Center,
Huazhong University of Science and Technology, Wuhan 430074,  China}
\author{X. C. Xie}
\affiliation{International Center for Quantum Materials, School of Physics, Peking University, Beijing 100871, China }
\affiliation{Collaborative Innovation Center of Quantum Matter, Beijing 100871, China}
\affiliation{CAS Center for Excellence in Topological Quantum Computation, University of Chinese Academy of Sciences, Beijing 100190, China}
\begin{abstract}
We study the electronic structure of a double-twisted few layer graphite (DTFLG), which consists of three few layer graphite (FLG), \textit{i.e.}~ABA-stacked graphene multilayer, stacked with two twist angles. We consider two categories of DTFLG, alternately twisted case  and chirally twisted one,  according to the rotation direction of the two twist angles. We show that, once the middle FLG of DTFLG is not thinner than trilayer, both kinds of  DTFLG can remarkably host two pairs of degenerate moir\'{e} flat bands (MFBs)  at $E_f$,  twice that of the magic angle twisted bilayer graphene (TBG).  The doubled MFBs of DTFLG lead to doubled DOS at $E_f$, which implies much stronger correlation effects than the TBG. The degeneracy of MFBs can be lifted by a perpendicular electric field,  and the isolated MFBs have nonzero valley Chern number. We also reveal the peculiar wave function patterns of the MFBs in the DTFLG. 
Our results establish a new family of moir\'{e} systems that have the much larger DOS at $E_f$, and thus possible much stronger correlation effects.  
\end{abstract}
\maketitle

\emph{Introduction.}---Twisted bilayer graphene (TBG) host a pair of moir\'{e} flat bands (MFBs) at the so-called magic angle (per valley per spin) \cite{mac2011,prl2007,prb2012,prb2012}.  The  giant  density  of states (DOS) and quenched kinetic energy of the MFBs are especially  conductive  to  interaction-driven  states.   Mott-like insulating state and unconventional superconductivity have already been observed in TBG, which has drawn great research interest very recently\cite{cao2018b,cao2018a,Jiang201991,Yankowitz2019,lau2019,Kere2019,koshino2018,kangjian2018}. 
Actually, MFB  is a general phenomenon of moir\'{e} heterostructures (MHSs). 
It exists in many other similar MHSs, such as twisted double bilayer graphene \cite{yhzhang2019,Koshino,jeil2019,lee2019,2019arXiv190308130L,cao2019,2019arXiv190308130L,DFTdoublebilayer,tutu2019,Wudoublebilayer,Samajdar2020,haddadi2020moire,zhangguangyu2019}, twisted trilayer graphene \cite{MA202118,kaxiras2019,chenshaowen2020,brey2013,shi2020,young2020,helin2018,Rademaker2020,jung2021_2},  trilayer graphene on boron nitride\cite{jeilprl2019,wangfeng2019,Chennature}, twisted few layer graphite\cite{jinhuaGao2020,zhang2020_1},  \textit{etc.}  Interestingly, the MFBs in some of these MHSs are topological nontrivial\cite{yhzhang2019,Koshino,jeil2019,lee2019,Jianpeng2019,MA202118,Fengcheng2019}, and topological phenomena, \textit{e.g.}~intrinsic quantum anomalous hall effect\cite{Serlin900,Lu2019653,Sharpe2019605},  can arise due to these topological MFBs. Very recently, MFBs in multi-twisted MHSs also have been studied \cite{Khalaf2019,lixiaotrilayer,Mora2020,zhu2020,Tritsaris2020,tritsaris2020lan,Zhu20202,Fengcheng2020,Zhu20203,cea2019twists,carr2020ultraheavy,tsai2019,Park2021}.


In general, the larger DOS is, the stronger electronic correlation is.
Thus, a fascinating question is  whether it is possible to further enhance the DOS of the moir\'{e} flat band systems, so that the correlation phenomena in MHSs can be more evident  with a higher transition temperature. At first glance, it is hardly possible since that a flat band already reaches the DOS maximum of a single band. 
Note that, in nearly all the MSHs reported so far, we can get at most  a pair of MFBs at the Fermi level $E_f$. 

\begin{figure}
\centering
\includegraphics[width=8cm]{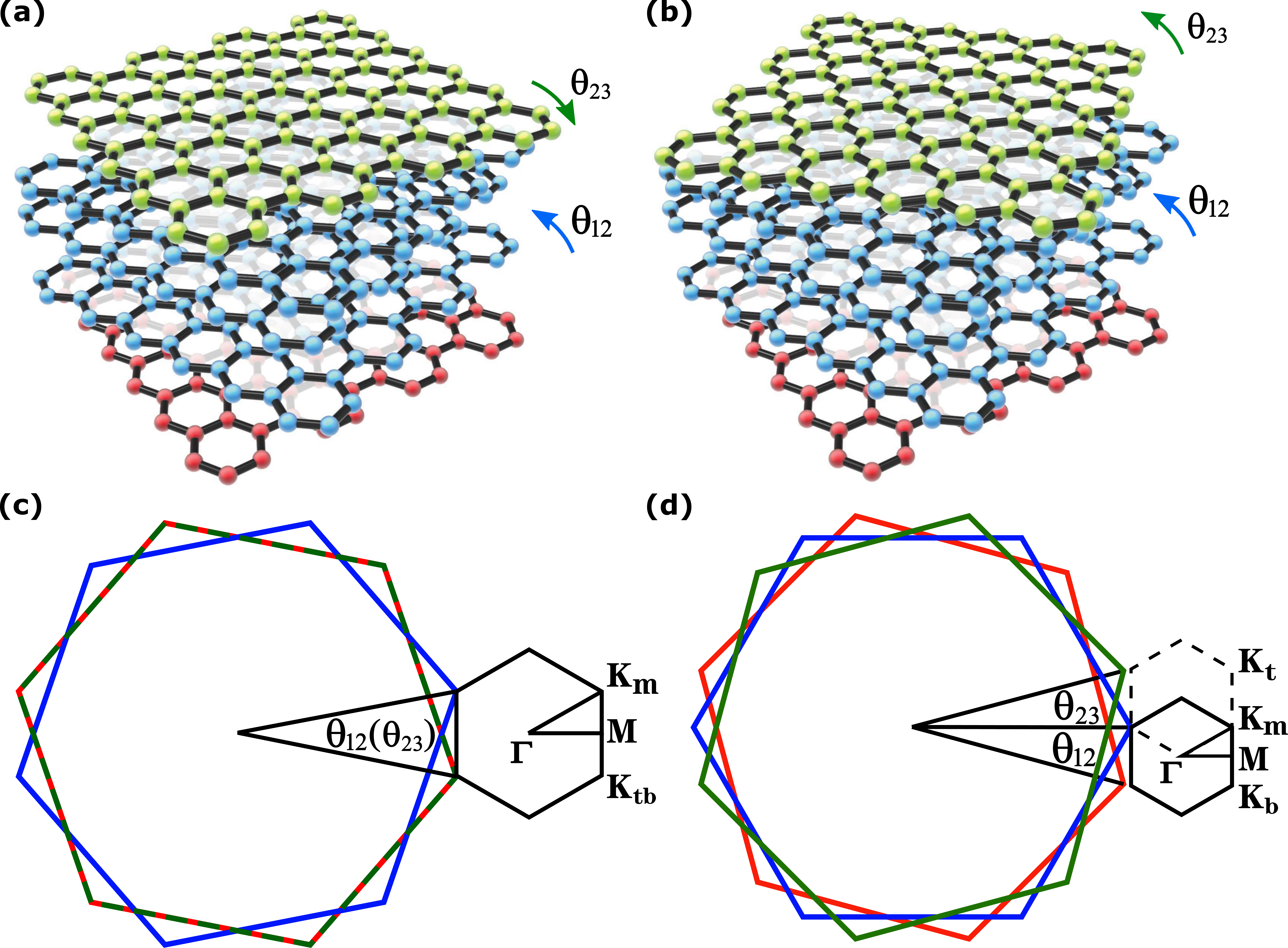}
\caption{(a) is the schematic of the (1+3+1)-aDTFLG and  (c) is the corresponding moir\'{e} BZ. (b) is the schematic of the (1+3+1)-cDTFLG and (d) is the corresponding moir\'{e} BZ.   $\theta_{12}$ and $\theta_{23}$ are the two twist angles. Red, blue and green represent the bottom, middle and top vdW layer, respectively.}
\label{fig1}
\end{figure}

\begin{figure*}
\centering
\includegraphics[width=17.5cm]{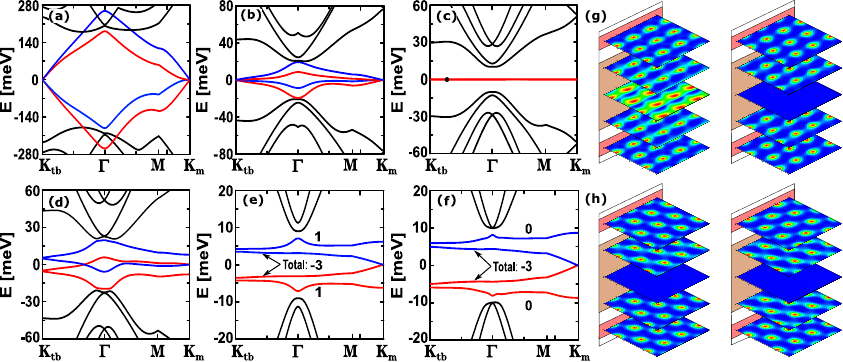}
\caption{ (a-f) are the moir\'{e} band structure of a (1+3+1)-aDTFLG. The parameters are: (a) $\theta=2.88^\circ$,  (b) $\theta=1.25^\circ$, (c)  $\theta=1.08^\circ$, (d) $\theta=1.25^\circ$ and $V=5$ meV, (e) $\theta=1.08^\circ$ and $V=5$ meV, (f)  $\theta=1.08^\circ$ and $V=7$ meV. 
(g) and (h) show the  flat band wave function distribution in each graphene monolayer, which correspond to the k state marked (black dot) in (c).  (g) is for 2nd MFB of the total 4 MFBs from bottom to up, and (f) is for the top MFB. Left (right) column is for A (B) sublattice of each graphene monolayer. The parameters:~$\omega_{\rm{AA}}=0.0797$~eV,$\omega_{\rm{AB}}=0.0975$~eV. }
\label{fig2}
\end{figure*}

 In this work,  we report a new MHS family, \emph{double-twisted few layer graphite} (DTFLG), which   remarkably can host \emph{two} pairs of  degenerate MFBs at the Fermi level, \textit{i.e.}~twice as much as  that in TBG.  
  Since the number of flat bands has doubled, the DOS of the DTFLG can be twice that of the magic angle TBG, which implies much  stronger correlation effects and possible higher transition temperature than the  magic angle TBG. 
 DTFLG is a sandwich-like MHS, which consists of three few layer graphite (FLG), \textit{i.e.}~ABA-stacked graphene multilayer,  stacked with two twist angles.
 We show that once the sandwiched middle FLG is not thinner than trilayer, no matter the DTFLG is alternately \cite{Khalaf2019} or chirally \cite{Mora2020} twisted,  the DTFLG always have two pairs of degenerate MFBs at certain magic angle. 
 Meanwhile, different band structures of MFBs, \textit{i.e.}~isolated flat bands or coexisting with other dispersive bands, can be achieved by choosing different stacking arrangements of DTFLG. Such MFBs in DTFLG are topological nontrivial and can be further adjusted by a perpendicular electric field. In experiment, the double twisted MHS  has already been realized in  recent experiments\cite{tsai2019,Park2021}, so that the  fabrication of DTFLG shoud be  not too challenging. Our study indicates that the DTFLG is a promising moir\'{e} platform that can realize stronger correlation  effects, and double twist can give rise to unique flat physics that is absent in single-twist MHSs like TBG.

\emph{Structure of DTFLG.}---A general (M+N+P)-DTFLG  is composed of three van der Waals (vdW) layers twisted relative to the adjacent layer, where each vdW layer is a FLG. Here, M, N and P represent the layer number of the bottom (1st vdW layer), middle (2nd vdW layer) and top FLG (3rd vdW layer), respectively.  Fig.~\ref{fig1} shows  the  typical  example of  (1+3+1)-DTFLG, in which we use  $\theta_{12}$ ($\theta_{23}$) to denote the twist angle between the middle  and bottom (top and middle) vdW layers.  Note that, according to sign of the two twist angles $\theta_{12}$ and $\theta_{23}$, we have two kinds of DTFLG: (1) alternately twisted DTFLG (aDTFLG) in Fig.~\ref{fig1} (a) with $\theta_{12} = -\theta_{23}$ \cite{Khalaf2019}; (2) chirally  twisted DTFLG (cDTFLG) in Fig.~\ref{fig1} (b) with $\theta_{12}=\theta_{23}$ \cite{Mora2020}.  Here, we always assume that $\theta \equiv |\theta_{12}|=|\theta_{23}|$, since otherwise the supercell of DTFLG does not exist. Note that, when $\theta=0$, both (1+3+1)-aDTFLG and (1+3+1)-cDTFLG return to a AABAA-stacked graphene quintuple-layer.  The moir\'{e} Brillouin zones (BZ) of the aDTFLG and cDTFLG are given in Fig.~\ref{fig1} (c) and (d), respectively. 

\begin{figure*}
\centering
\includegraphics[width=17cm]{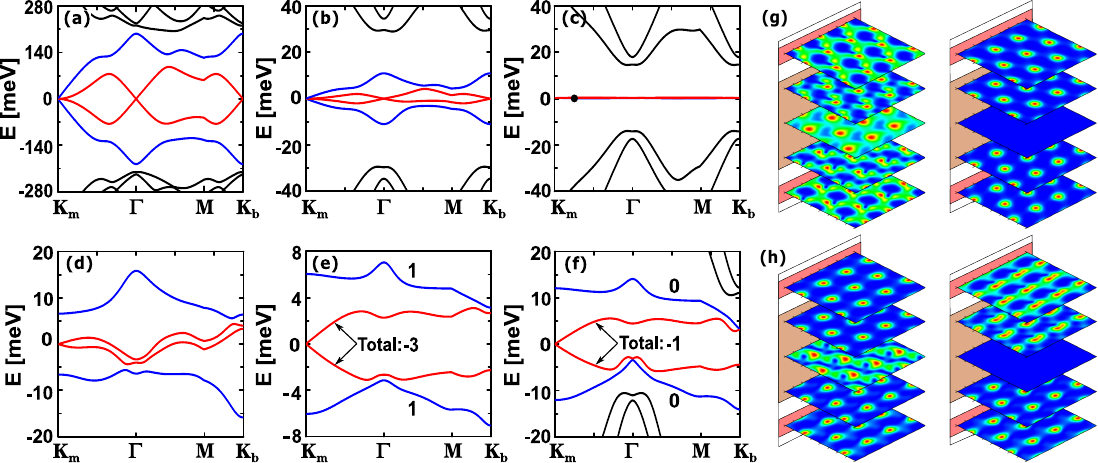}
\caption{(a-f) are the moir\'{e} band structure of a (1+3+1)-cDTFLG. The parameters are: (a) $\theta=2.88^\circ$,  (b) $\theta=1.25^\circ$, (c)  $\theta=1.08^\circ$, (d) $\theta=1.25^\circ$ and $V=5$ meV, (e) $\theta=1.08^\circ$ and $V=5$ meV, (f)  $\theta=1.08^\circ$ and $V=10$ meV. 
(g) and (h) show the  flat band wave function distribution in each graphene monolayer, which correspond to the k state marked (black dot) in (c).  (g) is for 2nd MFB of the total 4 MFBs from bottom to up, and (f) is for the top MFB. Left (right) column is for A (B) sublattice of each graphene monolayer. The parameters:~$\omega_{\rm{AA}}=0.0797$~eV, $\omega_{\rm{AB}}=0.0975$~eV. }
\label{fig3}
\end{figure*}

Not all the DTFLG has two pairs of MFBs. The key requirement is that $N \geq 3$, \textit{i.e.}~the middle FLG at least should be a ABA-stacked graphene trilayer. So, the (1+3+1)-DTFLG is the simplest case, which has two degenerate MFBs.  The reason is because the layer number of FLG determines the number of energy bands at $E_f$. 
A N-layer FLG has N (N-1)  parabolic bands when N is even (odd)  near the $E_f$, together with an additional pair of linear bands if N is odd. Therefore, ABA-trilayer graphene has two pairs of energy bands at $E_f$. Interestingly, when such two pair of energy bands are coupled with the  top and bottom  vdW layers via moir\'{e} interlayer hopping, they are simultaneously flattened at certain magic angle, so that we finally get  two pair of degenerate MFBs at  zero energy.  This novel phenomenon exist in all the DTFLG with $N \geq 3$ (both alternately twist and chirally twist cases), which is the main result of this work.


\emph{Continuum model.}---We use continuum model to calculate the moir\'{e} band structure of the DTFLG. Both aDTFLG and cDTFLG can be described by an unified Hamiltonian, which reads as a $3 \times 3$ block matrix 
\begin{equation}\label{eq1}
H_{\rm{DTFLG}}=\left( \begin{array}{ccc}
H_{1}(k_1)& T_{12}(r)& 0 \\
T_{12}^\dagger(r)&H_{2}(k_2)& T_{23}(r)\\
0&T_{23}^\dagger(r)&H_{3}(k_3)
\end{array} \right).
\end{equation}
The diagonal matrices are the tight-binding Hamiltonian of FLG in the Bloch wave basis, and the off-diagonal matrices describe moir\'{e} interlayer hopping. For example, $H_{1}(k_1)$ describes the bottom vdW layer, \textit{i.e.}~the M-layer FLG in a general (M+N+P)-DTFLG, which is a  $2M \times 2M$ matrix. $k_1=k-K_1$, where $k$ is the momentum relative to the $\Gamma$ point and $K_1$ is the Dirac point of the bottom vdW layer. $H_2(k_2)$ and $H_3(k_3)$ are given in the same way, which describe the middle and top vdW layer, respectively.   
The moir\'{e} interlayer hopping between adjacent vdW layers are 
$T_{ij}(r)=\sum_{n=0,1,2}T_{ij}^n\cdot e^{-iq_n\cdot r}$,  where $q_n=2k_D\sin(\frac{\theta}{2})\exp({i\frac{2n\pi}{3}})$ and  $k_D$ is the magnitude of the BZ corner wave vector of a single  vdW layer. And,
\begin{equation}\label{H1}
T_{ij}^{n}= I_{ij}\otimes\left( \begin{array}{ccc}
\omega_{\rm{AA}}&\omega_{\rm{AB}}e^{i\phi_n}\\
\omega_{\rm{AB}}e^{-i\phi_n}&\omega_{\rm{AA}}
\end{array} \right).
\end{equation}
Here,$I_{ij}$ is a matrix with only one nonzero matrix element. For instance, in a (M+N+P)-DTFLG, the only nonzero matrix element of $I_{12}$  is $I_{12} (M,N)=1$.  $\phi_n=\rm{sign}(\theta_{ij}) \frac{2n\pi}{3}$, in which the sign of $\theta_{ij}$ denotes the rotation direction. Note that  the above formulas are only valid for the cases with a small twist angle $\theta$.

\emph{Flat bands of (1+3+1)-aDTFLG.}---We first discuss the moir\'{e} band structure of the (1+3+1)-aDTFLG, as an example of the aDTFLG. 
In Fig.~\ref{fig2} (a), we plot the moir\'{e} band with a large angle $\theta=2.88^\circ$. Here, $K_m$  belongs to the middle FLG (ABA-trilayer), and $K_{tb}$  is from the top and bottom FLGs (graphene monolayers). Without moir\'{e} interlayer hopping, near the $K_{m}$, there are a pair of linear bands and a pair of parabolic bands from the middle ABA-trilayer. Meanwhile, the top and bottom graphene monolayer each provides one pairs of linear bands at  $K_{tb}$  with the same Fermi velocity.  At large $\theta$, we can see that the moir\'{e} interlayer hopping hybridizes the two pairs of bands at  $K_{m}$ and $K_{tb}$, and finally  two pairs of mori\'{e} bands are formed near the charge neutrality point, see the blue and red solid lines in Fig.~\ref{fig2} (a). Interestingly, due to the moir\'{e} interlayer hopping, the Fermi velocity of the two pairs of  linear bands at $K_{tb}$ becomes different, so that they are not degenerate any more.  
Decreasing $\theta$, the band width of the four moir\'{e} bands are narrowed. 
Around $\theta=1.25^\circ$, the two pair of central moir\'{e} bands  are separated from other high energy bands and a small gap appears, see Fig.~\ref{fig2} (b). Remarkably, when $\theta$ approaches $1.08^\circ$, the two pairs of  moir\'{e} bands  becomes   complete flat  (bandwidth is smaller than 1 meV) almost at the same time, see Fig.~\ref{fig2} (c). These four MFBs at zero energy are separated from other high energy bands by a gap about 10 meV.  

We emphasize that  the  double twist structure and the multi-bands feature of the FLG ($N\geq3$) are the two indispensable requirements for obtaining the two pairs MFBs. Here, we give two counterexamples.  One is the (1+1+1)-aDTFLG (\textit{i.e.}~double-twisted trilayer graphene), which do not satisfy the demand for $N\geq3$. Since the middle vdW layer just has two bands at $E_f$, the (1+1+1)-aDTFLG can only host a pair of MFBs, coexisting with a pair of linear bands\cite{Khalaf2019}. The other is the  (ABA+ABA)-type twisted FLG, which is  a single twist MHS. Though  it also has four bands near  $E_f$ at the two inequivalent Dirac points of the moir\'{e} BZ, a single twist can only induce one pair of MFBs at zero energy as well~\cite{jinhuaGao2020}. Our calculations indicate that  all the DTFLG with $N \geq 3$, meeting these two requirements above,   have two pairs of degenerate MFBs at the magic angle about $1.08^\circ$, the magic angle of which is the same as that of TBG but different from that of the double twisted trilayer graphene (about $1.5^\circ$)\cite{Khalaf2019}.


We can use a perpendicular electric field to lift the degeneracy of the flat bands.
Note that flat band states at $K_{tb}$ point come from the top and bottom vdW layers.  If a perpendicular electric field is applied, each vdW layer in the DTFLG feels different potential, and thus the degeneracy of the flat bands is lift. 
Here, we use the potential difference between the adjacent layers $V$ to represent the electric field. In Fig.~\ref{fig2} (d), we apply an electric field $V=5$ meV ($\theta=2.88^\circ$). We see that,  at $K_{tb}$, two moir\'{e} bands (blue lines) are shifted  upwards, which implies that they are mainly from the bottom graphene monolayer. Similarly, the other two bands shifted downwards (red lines) are from the top layer.  At $K_m$,  the four bands near $E_f$ are from the middle ABA-trilayer. Under the electric field, the two outer bands are gapped, while the inner two are still degenerate at $K_m$. Such band behaviours under electric field is in agreement with that of a ABA-trilayer. Similar phenomenon can be observed at the magic angle, see Fig.~\ref{fig2} (e) and (f).


The moir\'{e} flat bands of the (1+3+1)-aDTFLG are topological nontrivial. Without electric field, the valley Chern number of the moir\'{e} flat bands are not well-defined,  since the moir\'{e} flat bands are all degenerate at zero energy. However, electric field can lift all the degeneracy of the four MFBs, except for the degeneracy of the middle two bands at $K_m$ point, see Fig.~\ref{fig2} (e).  Note that there is a tiny gap between the top (bottom) two  MFBs at $K_{tb}$, which can be further enlarged by increasing the electric field, see Fig.~\ref{fig2} (f). Thus, the outer two MFBs are isolated. We then calculate the valley Chern number of the two isolated MFBs with the standard formula. When $V=5$ meV, the valley Chern numbers of the two moir\'{e} flat bands  are 1, see Fig.~\ref{fig2} (e).  And, though the middle two MFBs are still connected,  the total valley Chern number of the two bands are 3, also a nonzero value. Such topological nontrivial  MFBs indicate that, once the valley degeneracy is lift by the symmetry broken, orbital magnetization, quantum anomalous hall effects and other valley Chern number related topological phenomena may occur, just like that in other topological moir\'{e} systems.  Meanwhile, the interplay between correlation and band topology in DTFLG should be different from former moir\'{e} systems, because of its multi-flat-band feature. 

The valley Chern number can be changed by adjusting the electric field. For example, if we apply a larger electric field $V=7$ meV, a band cross can happen between the top (bottom) MFB and higher excited bands. As shown in Fig.~\ref{fig2} (f), a change of valley Chern number from 1 to 0 occurs simultaneously in the outer two isolated MFBs.

Typical wave functions of the MFBs in (1+3+1)-aDTFLG  are shown  in Fig.~\ref{fig2} (g) and (h), which corresponds to the denoted k point in Fig.~\ref{fig2} (c). Fig.~\ref{fig2} (g) is for the 2nd MFB from bottom to top, where the left (right) column is the wave function at sublattice A (B) in each graphene monolayer.  Interestingly, we find that over $75\%$ of the wave function lie in the sublattice A of the middle graphene monolayer, while that is zero in the sublattice B of the middle graphene monolayer, see Fig.~\ref{fig2} (g). Fig.~\ref{fig2} (h) is for the top MFB, where the wave function is zero at the middle graphene monolayer. The wave function of the bottom MFB is like the top one, while that of the middle two MFBs are similar.

\emph{Flat bands in (1+3+1)-cDTFLG.}---We then discuss the moir\'{e} band structures of the (1+3+1)-cDTFLG. The moir\'{e} BZ of the cDTFLG is different from that of aDTFLG. As shown in Fig.~\ref{fig1} (d).   The Dirac cones at zero energy of the top, middle and bottom vdW layers are attached to the $\Gamma$, $K_m$, and $K_b$ points, respectively. 
 Electronic states around the three Dirac points are coupled by the moir\'{e} interlayer hopping to form the moir\'{e} band structures at $E_f$.  

\begin{figure}
\centering
\includegraphics[width=8.5cm]{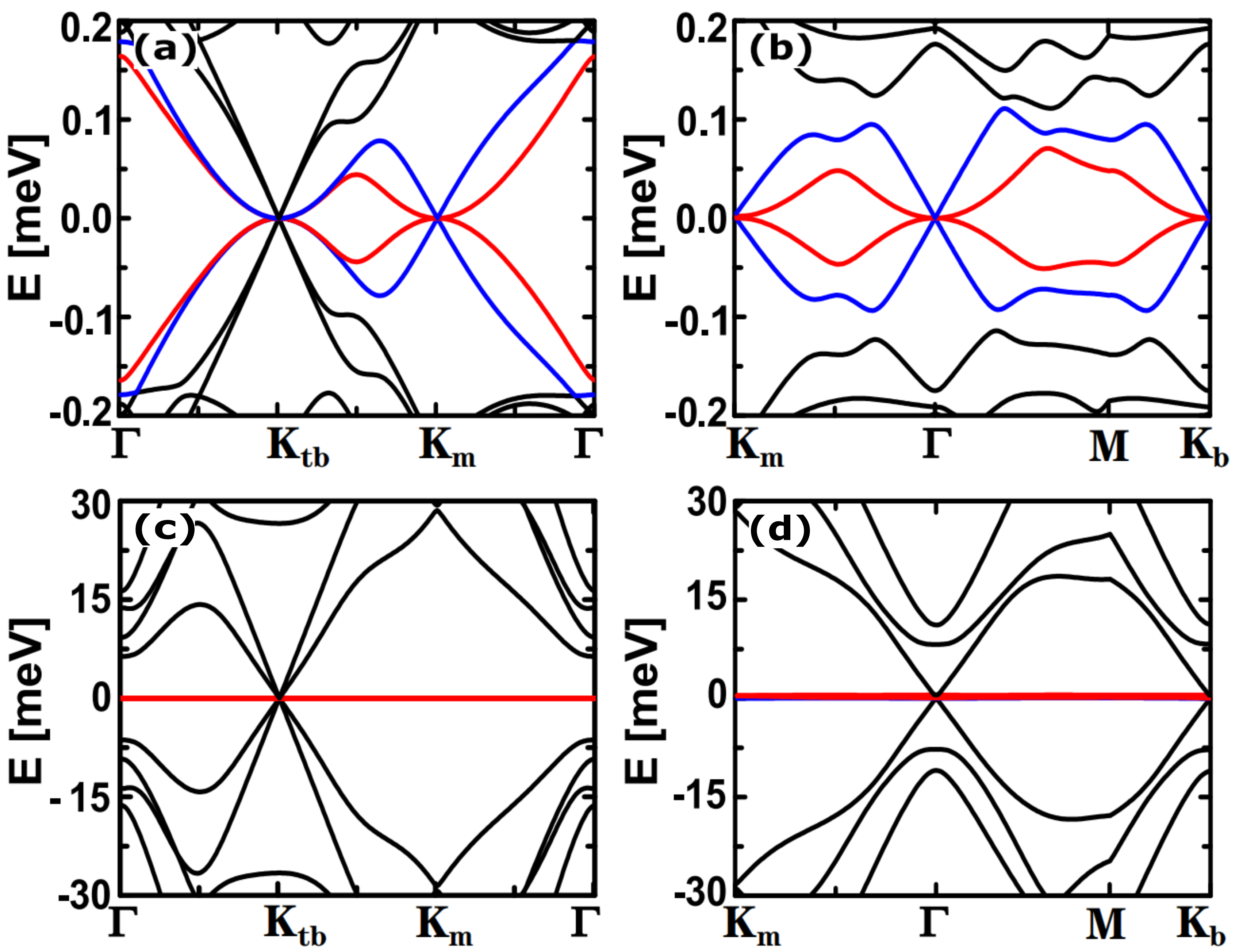}
\caption{ Moir\'{e} band structures of the (3+3+3)-DTFLG. (a) and (c) are for the alternately twisted case with $\theta=2.88^\circ$ and $\theta=1.08^\circ$, respectively. (b) and (d) are for the chirally twisted case with $\theta=2.88^\circ$ and $\theta=1.08^\circ$, respectively.}
\label{fig4}
\end{figure}

In the chirally twisted case, (1+3+1)-cDTFLG also has fundamentally different moir\'{e} band structure from that of (1+1+1)-cDTFLG (\textit{i.e.}~ a double twisted trilayer graphene), though their moir\'{e} BZs are similar.  
The (1+1+1)-cDTFLG is a perfect metal, where the moir\'{e} bands are always connected and gapless at all energy. And it does not have perfect flat bands. In contrast, as shown in Fig.~\ref{fig3}, the (1+3+1)-cDTFLG has  four moir\'{e} bands at  zero energy, gapped from other excited bands. Fig.~\ref{fig3} (a) gives the moir\'{e} bands with a large twist angle, $\theta=2.88^\circ$. We see that there are a pair of linear bands (blue line) and a pairs of parabolic bands (red  lines) around $K_m$, resulted from middle vdW layer. And,  there are only two linear bands (red solid line) near $E_f$ at either $\Gamma$ (from top vdW layer) or $K_b$ (from bottom vdW layer) points. 
 When $\theta$ is reduced, the band width of the four moir\'{e} bands are narrowed, and the gap between flat bands and dispersive bands becomes obvious, see Fig.~\ref{fig3} (b). At the magic angle $\theta=1.08^\circ$, the band width becomes nearly zero, and we get four degenerate MFBs, see Fig.~\ref{fig3} (c).

Here, the degeneracy of the four  MFBs can be lift by a perpendicular electric field as well, as shown in Fig.~\ref{fig3} (d), (e) and (f) with different $\theta$ and $V$. The outer two MFBs (blue lines) are isolated by the electric field, while the middle two (red) are still connected at $K_m$.  We further calculate the valley Chern number of the four MFBs, as denoted in Fig.~\ref{fig3} (e) and (f). The valley Chern number depends on the electric field. When we increase V from 5 meV to 7 meV,  a Chern number +1 is transferred from outer two MFBs to the central two. 


The wave functions of the outer (inner) two MFBs are similar.
Fig.~\ref{fig3} (g) gives the wave function of  the 2nd MFB from bottom to up,
while Fig.~\ref{fig3} (f) shows that of the top MFB. In the two situations, the wave function on the B sublattice of the middle graphene monolayer is always zero. Meanwhile, for the inner two bands, see Fig.~\ref{fig3} (g), rather large part of the wave function (about $53\%$) lies in the A sublattice of the middle graphene monolayer. As for the outer two MFBs, the corresponding value is only $18\%$.

\emph{Flat bands in (3+3+3)-DTFLG.}---In the (1+3+1)-DTFLG cases above, we always get two pairs of MFBs at $E_f$ gapped from other excited bands. If we use thicker FLG to build the DTFLG, we can introduce additional dispersive bands at $E_f$, coexisting with the two pairs of MFBs. As an example, in Fig.~\ref{fig4}, we plot the moir\'{e} bands of (3+3+3)-DTFLG, where each vdW layer is a ABA-trilayer graphene. Fig.~\ref{fig4} (a), (c) are  the alternating twisted configuration with $\theta=2.88^\circ$ and $\theta=1.08^\circ$, respectively. We see that there are eight bands at $E_f$, the band width of which are all narrowed with decreased $\theta$. At the magic angle $\theta=1.08^\circ$, we get not only  two pairs of MFBs but also two pairs of linear bands with different Fermi velocity at $K_{tb}$. Fig.~\ref{fig4} (b), (d) are the chirally twisted configuration accordingly. At large $\theta$, there are four bands at $E_f$, see Fig.~\ref{fig4} (b). But when decreasing $\theta$, two additional bands quickly get close to the $E_f$ near   $\Gamma$ and $K_b$, and at $\theta=1.08^\circ$ the middle four bands becomes completely flat, see Fig.~\ref{fig4} (d).

\emph{Summary.}---We have shown that a DTFLG, both the alternately and chirally twist situations,  can have two pairs of MFBs at $E_f$, twice that of the magic angle TBG. Thus, it gives rise to a doubled DOS at $E_f$, which implies stronger correlation effects than all the known MHSs. Such MFBs in DTFLG are also topolgical nontrivial, and thus offers an promising platform to study the interplay between the correlation and band topology. 
A very notable issue is that the (1+3+1)-aDTFLG and (1+3+1)-cDTFLG has similar flat band structure but different crystal symmetry and MFB wave functions. So, an interesting question is that if such distinction (symmetry and wave functions) can lead to different correlation phases.  In experiment,  double twisted trilayer graphene has already been fabricated very recently. Thus, we believe that our prediction can be readily tested in future experiment.  Especially, we argue that the predicted two pairs of MFBs in DTFLG can be directly observed by the nanoAPRES technique.

\begin{acknowledgments}
We thank the supports by the National Natural Science Foundation of China (Grants No.11874160, 11534001, 11274129, 11874026， 61405067), and the Fundamental Research Funds for the Central Universities (HUST: 2017KFYXJJ027), and NBRPC (Grants No. 2015CB921102). 
\end{acknowledgments}

\bibliography{twistedgraphene}
\end{document}